\documentclass[%
 reprint,
nofootinbib,
 amsmath,amssymb,
 aps,
 prl,
longbibliography,
]{revtex4-2}

\usepackage{graphicx}
\usepackage{dcolumn}
\usepackage{bm}
\usepackage{hyperref}
\usepackage{tikz}
\usetikzlibrary{decorations.markings,decorations.pathmorphing}
\usetikzlibrary{angles,quotes} 
\usetikzlibrary{arrows.meta} 
\colorlet{myred}{red!100}
\tikzstyle{softgraviton}=[-{Latex[length=4,width=3]},myred,line width=0.4,decorate,
                    decoration={snake,amplitude=0.9,segment length=4,post length=3.8}]

\begin{document}

\preprint{APS/123-QED}

\title{Counting Black Hole Microstates at Future Null Infinity}

\author{Vahid Reza Shajiee}
 \affiliation{%
 School of Physics, Institute for Research in Fundamental
Sciences (IPM), P.O.Box 19395-5531, Tehran, Iran
}
 \email{v.shajiee@ipm.ir}

\date{\today}

\begin{abstract}
We propose a black hole microstate counting method based on the canonical quantization of the asymptotic symmetries of a two-dimensional dilaton-gravity system at future null infinity. This dilaton-gravity is obtained from the s-wave reduction of an $N$-dimensional Einstein gravity over a generic asymptotically flat black hole solution.  We show that a Cardy-type formula leads to the Bekenstein-Hawking entropy. In this scenario, the quantized bulk degrees of freedom act as a source for the dual field theory operators living at future null infinity that are soft gravitons of the higher dimensional system, up to some gauge fixing.

\end{abstract}

\maketitle


\section{\label{sec1}Introduction}

Classically, information inside black holes is not accessible to observers outside of them, so they can be viewed as thermodynamic systems with temperature and entropy which are governed by thermodynamic laws~\cite{Bekenstein:1972tm,Bekenstein:1973ur}. Accounting for quantum effects led to the discovery of Hawking radiation~\cite{Hawking:1974rv,Hawking:1975vcx}, which raised concerns that black holes might violate the unitarity principle~\cite{Hawking:1976ra}. The emergence of holographic ideas~\cite{tHooft:1993dmi,Susskind:1994vu}, especially the Anti-de Sitter/Conformal Field Theory (AdS/CFT) correspondence~\cite{Maldacena:1997re,Aharony:1999ti} which provides concrete and practical calculations for the idea of holography, gave hope that information may somehow be read off from the boundary of quantum gravity spacetime. Black holes are expected to also have a microscopic description as thermodynamic systems; therefore, counting black hole microstates should match the Bekenstein-Hawking entropy, i.e $\text{S}_{\text{mic}}=\text{A}_{\text{hor}}/\text{4G}=\text{S}_{\text{BH}}$. Thanks to the holographic principle, since the maximum information storage capacity of black holes is equal to their entropy~\cite{Bekenstein:2001qi}, then counting black hole microstates plays a significant role in resolving the black hole information paradox mentioned above. We refer the reader to~\cite{Penington:2019kki,Balasubramanian:2022gmo,Balasubramanian:2022lnw,Boruch:2023trc,Climent:2024trz} for recent studies on the black hole information paradox and the microstate counting program.

 Using the Cardy formula governing the asymptotic behavior of the density of states of a CFT~\cite{Cardy:1986ie}, numerous papers achieved the goal of successfully counting microstates of string-theoretic black holes~\cite{Strominger:1996sh,Birmingham:1998jt}, pure BTZ black holes~\cite{Strominger:1997eq}, generic black hole spacetimes~\cite{Carlip:1998wz,Solodukhin:1998tc}, and celestial black holes including extremal~\cite{Guica:2008mu} and near extremal cases\footnote{See~\cite{Compere:2012jk} and references therein.}. Motivated by the search for a holographic description of asymptotically flat black holes, including non-extremals, Barnich et al.~\cite{Barnich:2009se} showed that the symmetry algebra should be the semidirect sum of supertranslations with infinitesimal local conformal transformations, derived from the well-known Bondi–Metzner–Sachs (BMS) group which is the group of asymptotic isometries at null infinity~\cite{Bondi:1962px,Sachs:1962wk}. By taking both the AdS radius in the bulk and the speed of light on the boundary to infinity, one recovers the $\text{BMS}_{3}/\text{GCA}_{2}$\footnote{Galilean Conformal Algebra (GCA)} correspondence from the $\text{AdS}_{3}/\text{CFT}_{2}$ correspondence~\cite{Bagchi:2010zz,Bagchi:2012xr}. Therefore, it is natural to seek a deformed CFT when the bulk spacetime deviates from AdS. Recently, using covariant phase space methods~\cite{Lee:1990nz,Carlip:1999cy,Barnich:2001jy} in two-dimensional (2D) dilaton gravity dimensionally reduced from higher-dimensional gravities, Carlip showed that the $\text{BMS}_{3}$ symmetry at the horizon may be responsible for the microscopic origin of the black hole entropy~\cite{Carlip:2017xne,Carlip:2019dbu}.

 According to the ``soft hair on black holes'' proposal, black holes may carry an infinite number of charges beyond the no-hair theorem~\cite{Hawking:2016msc}. The soft hairs allow for charge conservation by relating the charges carried by emitted or absorbed particles in every scattering process in the asymptotic region of spacetime to themselves~\cite{Hawking:2016sgy}. Consequently, the soft hairs follow the symmetry of the BMS subgroup.

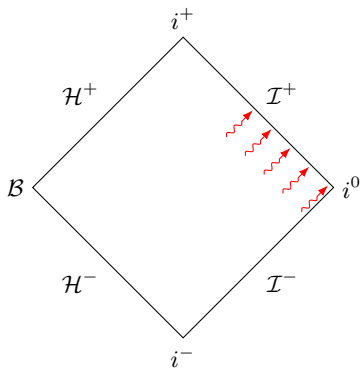
\begin{figure}
\centering
\begin{tikzpicture}[scale=0.5]

\node (I)   at (0,0)   {};

\path  
  (I)  +(90:4)  coordinate[label=90:$i^+$]  (Itop)
       +(-90:4) coordinate[label=-90:$i^-$] (Ibot)
       +(0:4)   coordinate[label=0:$i^0$] (Iright)
       +(180:4) coordinate[label=180:$\cal{B}$] (Ileft)
       ;
\draw (Ileft) --
          node[midway, above left]    {$\cal{H}^+$}
      (Itop) --
          node[midway, above right] {$\cal{I}^+$}
      (Iright) --
          node[midway, below right] {$\cal{I}^-$}
      (Ibot) --
          node[midway, below left]    {$\cal{H}^-$}
      (Ileft) -- cycle;
  \draw[softgraviton] (1.5,1.7) ++ (45:-0.5) --++ (45:1.0);
  \draw[softgraviton] (2.0,1.2) ++ (45:-0.5) --++ (45:1.0);
  \draw[softgraviton] (2.5,0.7) ++ (45:-0.5) --++ (45:1.0);
  \draw[softgraviton] (3.0,0.2) ++ (45:-0.5) --++ (45:1.0);
  \draw[softgraviton] (3.5,-0.3) ++ (45:-0.5) --++ (45:1.0);
\end{tikzpicture}
\caption{Penrose diagram of the exterior region of a black hole, including the representation of soft gravitons figuratively\label{fig1}}
\end{figure}

In the present work, we employ covariant phase space methods to compute the charges of the asymptotically flat black hole at future null infinity $\mathcal{I}^{+}$. We then utilize the Cardy formula to demonstrate that counting microstates of the dual field theory at $\mathcal{I}^{+}$ precisely reproduces the Bekenstein-Hawking entropy. A convincing reason to seek the dual field theory at $\mathcal{I}^{+}$ is that the soft gravitons in the higher dimensional system (formulated in~\cite{Weinberg:1965nx} and linked to BMS supertranslations in ~\cite{He:2014laa}) may bring properties of the dual field theory from the past horizon $\cal{H}^-$ to $\mathcal{I}^{+}$. In such a scenario for the higher dimensional system, the soft gravitons act as the sources for the operators of the deformed CFT at $\mathcal{I}^{+}$. So, when one performs a dimensional reduction, it is expected that the same physics governs the dilaton-gravity system, in which the dilaton field plays a key role in the absence of (soft) graviton, up to some gauge fixing. The Penrose diagram of the exterior region of a black hole, including the representation of soft gravitons, is shown in figure \ref{fig1}. As a reference to validate our findings, we employ Carlip's approach worked out in~\cite{Carlip:2017xne,Carlip:2019dbu}.

\section{\label{sec2}Gravity setup}

When one studies (quantum) gravity in higher dimensions $N\geq4$, the physics usually gets complicated due to extra degrees of freedom admitting gravitons to propagate. On the other hand, there are pieces of evidence, from string theory\footnote{See~\cite{Klebanov:1991qa} and references therein.} to black hole physics~\cite{Frolov:1992xx,Banados:1993qp}, that two dimensional ``$r$--$t$'' plane is rich enough to handle the physics in question; nevertheless, if we want to model a higher dimensional gravity by 2D one, then the 2D model should have some extra fields to compensate, up to some gauge fixing, ``$N$$-$$2$'' reduced degrees of freedom. This can naturally be done with a dimensional reduction. The result, up to an integration over transverse coordinates, is the 2D dilaton gravity with boundary
\begin{align}
I = \frac{1}{16\pi G}\int_M\!\left(\varphi R + V[\varphi] \right)\epsilon + \frac{1}{8\pi G}\int_{\partial M}\!\varphi\left(K-K_{0} \right)\epsilon_{\partial M} \, ,
\label{gsIaction}
\end{align}
where $\epsilon$ and $\epsilon_{\partial M}$ are the bulk volume two-form and the boundary volume one-form, respectively. Because we are interested in the asymptotic region of spacetime, the Gibbons–Hawking–York boundary term is taken into account in action \ref{gsIaction} to have a well-posed variational principle. Here, the boundary includes both $i^0$ and $\mathcal{I}^{+}$. One approach to work with the null boundary is to define a projector that projects spacetime objects onto the co-dimension two surface~\cite{Jafari:2019bpw}. However, such a projector is identically zero in two dimensions. Therefore, it is expected that the term involving the null boundary does not alter calculations in the covariant phase space methods for two dimensions. From now on, we only work with the bulk term to avoid redundancy in calculations, nonetheless, we hold in mind that the term involving the spatial boundary may contribute to calculations. Variation of the action \ref{gsIaction} with respect to metric $g_{\mu \nu}$ and dilaton $\varphi$ fields respectively yields the equations of motion
\begin{subequations}
\begin{align}
&E^{\mu\nu} = \nabla^{\mu}\nabla^{\nu}\varphi - g^{\mu\nu}\Box\varphi + \frac{1}{2}g^{\mu\nu}V = 0 \, ,
   \label{gsEOMa}\\
&R + \frac{dV}{d\varphi} = 0 \, . \label{gsEOb}
\end{align}
\end{subequations}
Soft gravitons are a type of gravitational radiation at $\mathcal{I}^{+}$ in the higher dimensional system. The Newman--Penrose formalism, developed in~\cite{Newman:1961qr}, offers a powerful approach to studying gravitational radiation. In this formalism, the metric and Levi-Civita tensor reads
\begin{subequations}
\begin{align}
&g_{\mu\nu} = - \ell_{\mu} n_{\nu} - n_{\mu}\ell_{\nu} \, ,\label{gsmetnepsa}\\
&\epsilon_{\mu\nu} = \ell_{\mu} n_{\nu} -  n_{\mu}\ell_{\nu} \, .\label{gsmetnepsb}
\end{align}
\end{subequations}
where $\ell$ and $n$ are null vectors, $\ell_{\rho} \ell^{\rho}=n_{\rho} n^{\rho}=0$, which follow $\ell_{\rho} n^{\rho}=-1$. It is conventional to define two directional derivatives
\begin{align}
D=\ell^{\mu} \nabla_{\mu} \, , \quad \Delta = n^{\mu} \nabla_{\mu} \, .
\label{gsDDderivs}
\end{align}
To consider the exterior of a black hole in the Newman--Penrose formalism, the spacetime must include a horizon. For our purposes, this can be done by finding a boundary condition that gives a null surface $\cal{H}^-$ which is a non-expanding past boundary of the spacetime. The dilaton field is the only nontrivial Jacobi field in 2D dilaton gravity, so letting $n^{\mu}$ be normal to $\cal{H}^-$ and using the expansion formula along $n^{\mu}$, one easily gets\footnote{Omitting the $\pm$ signs, we show the equality on $\mathcal{I}^+$ and $\cal{H}^-$ by $\overset{\mathcal{I}}{=}$ and $\overset{\cal{H}}{=}$, respectively.}
\begin{align}
\theta\overset{\cal{H}}{=}\varphi^{-1}\Delta\varphi.
\label{gsExpan}
\end{align}
Clearly, the boundary condition $\Delta\varphi\overset{\cal{H}}{=}0$ gives the desired non-expanding horizon. In the presence of the horizon, the spacetime is equipped with a surface gravity $\kappa$ which gives the horizon surface gravity on $\cal{H}^-$. Thus, $n^{\mu}$ satisfies the geodesic equation $n^{\mu}\nabla_{\mu}n^{\nu} = \kappa n^{\nu}$, fixing the other null vector to preserve the same length while parallelly propagates along itself (i.e. $l^{\mu}\nabla_{\mu}l^{\nu} = 0$). These geodesic equations give
\begin{alignat}{3}
&\nabla_{\mu}\ell_{\nu} = \kappa \ell_{\mu}\ell_{\nu} \, , \qquad\qquad
   && \nabla_{\mu}\ell^{\mu} = 0 \, , \nonumber\\
&\nabla_{\mu} n_{\nu} =  -\kappa \ell_{\mu} n_{\nu} \, , && \nabla_{\mu} n^{\mu} = \kappa \, ,
\label{gsGeode}
\end{alignat}
The variation of these equations yields
\begin{subequations}
\begin{align}
&D(n^{\rho}\delta \ell_{\rho}) = (\Delta+\kappa)(\ell^{\rho}\delta \ell_{\rho}) \, ,\label{gsVarklna}\\
&\delta\kappa = D(n^{\rho}\delta n_{\rho}) +  \kappa n^{\rho}\delta \ell_{\rho}
   -  {\Delta}(\ell^{\rho}\delta n_{\rho}) \, .
\label{gsVarklnb}
\end{align}
\end{subequations}
The following identities will be used in later calculations,
\begin{subequations}
\begin{align}
&[\Delta,D] = -\kappa D\, ,
   \label{gsIdenta}\\
&R = 2D\kappa \, , \label{gsIdentb}\\
&(d\psi)_{\mu} = -D\psi\,n_{\mu} - \Delta\psi\,\ell_{\mu} \, . \label{gsIdentc}
\end{align}
\end{subequations}
The first identity is straightforward. The second one is easily obtained by applying Ricci identity for $n^{\mu}$. Thanks to the geodesic equations, the third one holds for any function $\psi$, where $n_{\mu}$ and $\ell_{\mu}$ are treated as one-forms~\cite{Carlip:2017xne}.

Note that we are dealing with a 2D dilaton gravity, so there is no soft graviton. We have just argued that the model is dimensionally reduced from a higher dimensional gravity, so the soft gravitons, discussed in the previous section and depicted in figure \ref{fig1}, should be perceived from the original theory. As we see later, the information that is carried by soft graviton is encoded in dilaton in two dimensions.

\section{\label{sec3}Covariant phase space formalism}

Local symmetries of Lagrangian formulation of field theories are in correspondence with constraints in phase space formulations, also, there is an isomorphism between
the Poisson bracket algebra of the constraints and the Lie bracket algebra of the local symmetries~\cite{Lee:1990nz}. So, a field theory problem can be turned into a phase space problem governed by Hamiltonian formulation in a covariant manner. The phase space is a symplectic manifold with a non-degenerate closed two-form given by
\begin{align}
\Omega[\Phi;\delta_1\Phi,\delta_2\Phi]
    &= \int_\Sigma \omega[\Phi;\delta_1\Phi,\delta_2\Phi]\nonumber\\
    &= \int_\Sigma \omega_{ij}\delta_1\Phi^i\wedge\delta_2\Phi^j \, ,
\label{cpsm-OmegaBig}
\end{align}
where $\Sigma$ is a Cauchy surface. $\Phi^i$ are the fields in Lagrangian, i.e. metric and dilaton in 2D dilaton gravity. $\omega$ is the symplectic current given by
\begin{align}
\omega[\Phi;\delta_1\Phi,\delta_2\Phi]
   &= \delta_1\Theta[\Phi,\delta_2\Phi] - \delta_2\Theta[\Phi,\delta_1\Phi]\, ,
   \label{cpsm-omegaSmall}
\end{align}
where $\Theta$ is the symplectic potential one-form. This boundary term can be obtained from integration by parts in the variation of the Lagrangian density, as given by
\begin{align}
\delta L = E_i\delta\Phi^i + d\Theta[\Phi,\delta\Phi] \, ,
\label{cpsm-varL}
\end{align}
where $E_i=0$ gives the equations of motion. The fields $\Phi^i$ and their variations $\delta\Phi$ are the classical solutions, respectively, to the equations of motion and the linearized ones. The symplectic two-form \ref{cpsm-OmegaBig} determines Hamiltonians and Poisson brackets for given canonical transformations $\delta_{\tau}\Phi^{i}$~\cite{Carlip:2019dbu}
\begin{subequations}
\begin{align}
\delta H[\tau] &= \Omega[\delta\Phi,\delta_\tau\Phi], \, \label{cpsm-dH}\\
\left\{ H[\tau_1],H[\tau_2]\right\} &= -\Omega[\delta_{\tau_1}\Phi,\delta_{\tau_2}\Phi]. \, \label{cpsm-PBra}
\end{align}
\end{subequations}

The bulk term of the action \ref{gsIaction} can be taken as $f(R)$ gravity with Lagrangian density $L=f(R)\epsilon/16\pi G=(\varphi R + V[\varphi])\epsilon/16\pi G$. The symplectic current reads~\cite{Bueno:2016ypa}\footnote{We have verified the results of~\cite{Bueno:2016ypa}.}
\begin{subequations}
\begin{equation}\label{cpsm-Allin1a}
\begin{aligned}
\omega &= f'(R) \omega_{\text{Ein}} + \frac{1}{16\pi G}  \epsilon_{\mu }  \Big [  \frac{1}{2} g^{\mu\beta} g^{\alpha\nu} g^{\rho\sigma}  \delta_1 g_{\rho\sigma} \delta_2 g_{\alpha\beta}  \nabla_\nu f'(R) \\
&+ \big (g^{\mu\beta} g^{\alpha\nu} - g^{\mu\nu} g^{\alpha\beta}   \big) \times \\
&\times \big(    \delta_1 ( f'(R)) \nabla_\nu \delta_2 g_{\alpha\beta} -  \delta_1 (\nabla_\nu f'(R) ) \delta_2 g_{\alpha\beta}  \big )   - [1 \leftrightarrow 2] \Big ] ,\\
\end{aligned}
\end{equation}
\begin{equation}\label{cpsm-Allin1b}
\omega_{\text{Ein}}  =  \epsilon_{\mu }S^{\mu\alpha\beta\nu\rho\sigma} \left ( \delta_1 g_{\rho\sigma} \nabla_\nu \delta_2 g_{\alpha\beta}  - \delta_2 g_{\rho\sigma} \nabla_\nu \delta_1 g_{\alpha\beta}   \right), \\
\end{equation}
\begin{equation}\label{cpsm-Allin1c}
\begin{aligned}
S^{\mu\alpha\beta\nu\rho\sigma} &= \frac{1}{16\pi G} \Big [ -  g^{\mu(\alpha} g^{\beta)(\rho} g^{\sigma)\nu} + \frac{1}{2} g^{\mu(\alpha} g^{\beta)\nu} g^{\rho\sigma}\\
& + \frac{1}{2} g^{\alpha\beta} g^{\mu(\rho}  g^{\sigma)\nu}     + \frac{1}{2} g^{\mu\nu} g^{\alpha(\rho} g^{\sigma)\beta}  -  \frac{1}{2} g^{\mu\nu} g^{\alpha\beta} g^{\rho\sigma} \Big],\\
\end{aligned}
\end{equation}
\end{subequations}
where $f'(R)=df(R)/dR=\varphi$, $\epsilon_{\mu} = \epsilon_{\mu\nu} dx^{\nu}$, and $\omega_{\text{Ein}}$ is the symplectic current for Einstein gravity $f(R)=R$.

Calculating the symplectic current is generally complicated. During deriving $\nabla_{\mu}\delta \ell_{\nu}$ and $\nabla_{\mu}\delta n_{\nu}$ from the equation \ref{gsGeode}, one encounters the variation of the Christoffel symbols $\delta\Gamma$. These terms give freedom to define $\nabla_{\mu}\delta \ell_{\nu}$ and $\nabla_{\mu}\delta n_{\nu}$ so that they can be used for specific purposes like some simplification, as they preserve the equation \ref{gsGeode}. The following definitions do the task,
\begin{subequations}
\begin{align}
\nabla_{\mu}\delta\ell_{\nu} &= \kappa \ell_{\mu} \delta\ell_{\nu} + \kappa \delta\ell_{\mu} \ell_{\nu} - \delta\kappa \ell_{\mu} \ell_{\nu} \, , \label{cpsm-CDdell}\\
\nabla_{\mu}\delta n_{\nu} &= -\kappa \ell_{\mu} \delta n_{\nu} - \kappa \delta\ell_{\mu} n_{\nu} - \delta\kappa \ell_{\mu} n_{\nu} \, , \label{cpsm-CDdeln}
\end{align}
\end{subequations}
they give $\omega_{\text{Ein}}=0$. So, we are left with the dilatonic part in the equation \ref{cpsm-Allin1a}.

\section{\label{sec4}Future null infinity and boundary conditions}

As argued before, we are interested in the future null infinity region $\cal{I}^+$. Hence, in the context of the symplectic form \ref{cpsm-OmegaBig}, a suitable Cauchy surface could be formed by combining the future horizon $\cal{H}^+$ and the future null infinity $\cal{I}^+$. As seen in figure \ref{fig1}, this surface would have endpoints at the bifurcation point $\cal{B}$ and spacelike infinity $i^{0}$. Since we are looking for canonical transformations that vanish everywhere except at the asymptotic region $\cal{I}^+$, we ignore the details of $\cal{H}^+$ and $\cal{B}$.

Based on the arguments presented above, our Cauchy surface coincides with the boundary of spacetime. Consequently, it is necessary to impose boundary conditions at $\cal{I}^+$. Our boundary conditions are:
\begin{enumerate}
  \item $\Delta\varphi\overset{\mathcal{I}}{=}(\kappa/2\pi)(\kappa\varphi^{N-2}a/(N-3))^{\frac{N-2}{N-3}}$ and $\Delta\Delta\varphi\overset{\mathcal{I}}{=}0$. Here, $a$ is just a number to distinguish information retrieved from Schwarzschild ($a$$=$$2$), extremal Kerr ($a$$=$$\sqrt{2}$) or extremal RN ($a$$=$$1$) black holes.  This boundary condition reminds us of the divergence-free condition $\mathcal{L}_{n}\varphi=\Delta\varphi\overset{\mathcal{I}}{=}0$~\cite{Ashtekar:1981bq,Ashtekar:1981sf} where in that case dilaton field would vanish at the asymptotic region in contrast with our work that it is divergent at $\mathcal{I}^{+}$. Note that $\Delta\varphi\overset{\mathcal{I}}{=}0$ as a boundary condition is too strong because it destroys any information encoded in the dilaton. It is not hard to see that in a coordinate system like Gaussian null, our fall-off condition is of order $\Delta\varphi\overset{\mathcal{I}}{=}\mathcal{O}(1/r^{N-2})$ for simplest cases like Schwarzschild or extremal RN, where $r$ is the usual radial coordinate. \label{FNIBC1}
  \item $\Delta R\overset{\mathcal{I}}{=}0$. This follows directly from the asymptotically flat condition $R_{\mu\nu}\overset{\mathcal{I}}{=}0$ for a given spacetime~\cite{Ashtekar:1981hw}. \label{FNIBC2}
  \item $\delta \ell_{\mu}\overset{\mathcal{I}}{=}0$. Similar to that of~\cite{Carlip:2017xne}, this can be viewed as a gauge-fixing condition. This is because there always exists a local Lorentz transformation to fix the integration measure $\ell_{\mu}$ at $\cal{I}^+$. If $\ell_{\mu}$ were not fixed, then there would be terms in which $\delta \ell_{\mu}$ appears as an integration measure. This condition only simplifies the calculations, but relaxing it brings complications to the calculations. \label{FNIBC3}
  \item $n^{\alpha}\delta n_{\alpha}\overset{\mathcal{I}}{=}0$. We impose this condition only to simplify the calculations. If the condition is relaxed, some terms would need to be carried through the calculations without any physical impact on the final results.\label{FNIBC4}
  \item $\nabla_{\mu}\delta\varphi=\delta\nabla_{\mu}\varphi\overset{\mathcal{I}}{=}-\ell^{\alpha}\delta n_{\alpha}\nabla_{\mu}\varphi$. Since this condition, like condition \ref{FNIBC3} and \ref{FNIBC4}, is intended solely for simplification, it can be relaxed. However, this relaxation would sacrifice the extreme simplicity and straightforwardness of the computations.\label{FNIBC5}
\end{enumerate}

These conditions simplify calculations drastically, the $\cal{I}^+$ segment of the symplectic two-form \ref{cpsm-OmegaBig} is then
\begin{align}
\Omega_{\cal{I}^+}[\Phi;\delta_1\Phi,\delta_2\Phi]
    = \frac{1}{8\pi G}\int_{\cal{I}^+} \left[ \delta_{1}\varphi\,\delta_{2}\kappa - \delta_{2}\varphi\,\delta_{1}\kappa \right] \ell_{\mu} \, .
\label{FNIOmegI}
\end{align}
where $\Phi\equiv(\varphi,g)$ and $\ell_{\mu}$ is treated as one-form. Although we have reached a simple and elegant formula, the variation of the dilaton field violates the boundary condition \ref{FNIBC1}. Put another way, $\delta(\Delta\varphi)$ leads to the non-integrability of charges. Fortunately, this can be cured using the same procedure presented in~\cite{Carlip:2017xne}. There always exists a diffeomorphism $\delta_{\zeta}$ which can cure non-integrability, if
\begin{align}
\delta(\Delta\varphi) + \zeta^{\mu}\nabla_{\mu}(\Delta\varphi) \overset{\mathcal{I}}{=} 0 \, .
\label{FNIDiffeoAFN}
\end{align}
Because the symplectic two-form is independent of the integration contour, a transverse diffeomorphism generated by $\zeta^{\mu}=\zeta \ell^{\mu}$ is a suitable choice, assuming the endpoints of the Cauchy surface remain fixed that is $\delta(\Delta\varphi)\overset{i^{0}}{=}0$  for the segment of interest. Using the boundary condition \ref{FNIBC4}, such a choice reads
\begin{align}
\zeta^{\mu} = \zeta \ell^{\mu} = -\frac{\Delta\delta\varphi}{{D}\Delta\varphi}\ell^{\mu} \, ,
\label{FNIDiffeElem}
\end{align}
where $\Delta\delta\varphi\overset{i^{0}}{=}0$. The procedure outlined above must be adopted to find a Hamiltonian that satisfies equation \ref{cpsm-dH}. In other words, the terms leading to the non-integrability, if any, must be removed from the variation of any Hamiltonian,
\begin{align}
 (\delta H)_{\text{Credible}} \equiv \delta H + \delta_{\zeta} H \, .
\label{FNIRealH}
\end{align}
Other potential solutions to the non-integrability problem for charges can be found in the literature, such as~\cite{Wald:1999wa,Barnich:2007bf,Donnelly:2016auv,Ruzziconi:2020wrb,Adami:2020ugu,Ciambelli:2021nmv,Geiller:2021vpg,Adami:2021nnf} and references therein. Recently, it has been proposed that assuming boundary fluctuations can make the charges integrable in the covariant phase space formalism~\cite{Adami:2024gdx,Golshani:2024fry}.

\section{\label{sec5}Symmetries at future null infinity}

As mentioned in section~\pageref{sec1}, the symmetry of asymptotically flat spacetimes at null infinity is the BMS group. Equivalently, the 2D dilaton gravity action \ref{gsIaction} is invariant under the diffeomorphisms generated by supertranslations $\xi^{\mu}=\xi n^{\mu}$ which are the generators of $\text{BMS}_{3}$ subalgebra. To ensure compliance with the boundary condition \ref{FNIBC3} and avoid violations, it is necessary to apply a Lorentz transformation, expressed as $\delta n^{\mu}=(\delta\lambda)n^{\mu}=(\Delta\xi)n^{\mu}$. This procedure yields
\begin{align}
&\delta_\xi n^{\mu} = 0 \, , \; &&\delta_\xi \ell^{\mu} = -\ell^{\mu} (\Delta+\kappa)\xi \, , \nonumber\\
&\delta_\xi g_{\mu\nu} = g_{\mu\nu}(\Delta+\kappa)\xi \, , \; &&\delta_\xi \varphi = \xi \Delta\varphi \, , \label{SaFNIdiffeos}
\end{align}
where $D\xi\overset{\mathcal{I}}{=}0$ is assumed.

There are other kinds of transformations that leave the action \ref{gsIaction} approximately invariant. One such approximate symmetry, introduced in~\cite{Carlip:2002be} as a near-horizon symmetry, is given by a specific shift in the dilaton field. Generally, the approximate symmetry is a symmetry of action near a region where the shift function is localized. We may linearly combine the shift transformation with the Weyl transformation as $\hat{\delta}_{\eta}\equiv \text{Weyl} + \text{Shift}$. Since any shift in the metric ($g_{\mu\nu}$) and the vectors ($\ell^{\mu}$,$n^{\mu}$) can always be gauged away, we may write
\begin{align}
\hat{\delta}_{\eta} \ell_{\mu} &= 0 \,, \qquad \hat{\delta}_{\eta} n_{\mu} = \hat{\delta}_{\eta}\varpi n_{\mu} \, , \nonumber\\
&\hat{\delta}_{\eta} g_{\mu\nu} = \hat{\delta}_{\eta}\varpi g_{\mu\nu} \, . \label{SaFNIWeylShiftVg}
\end{align}
The first one is imposed by hand in compatibility with the boundary condition \ref{FNIBC3}. For the dilaton field, we may write
\begin{align}
\hat{\delta}_{\eta} \varphi = \frac{\mathcal{N}-2}{2} \hat{\delta}_{\eta}\varpi\varphi + (\Delta+\kappa)\eta  \, . \label{SaFNIWeylShiftD}
\end{align}
where $D\eta\overset{\mathcal{I}}{=}0$ is assumed. Obviously, we are left with the shift symmetry for the dilaton field in two dimensions $\mathcal{N}$=2. Using \ref{gsIdentb}, \ref{gsVarklnb} and \ref{gsIdenta} yields ${\hat\delta}_{\eta}R = 2(\Delta+\kappa)D{\hat\delta}_{\eta}\varpi$. Then, defining
\begin{align}
{\hat\delta}_{\eta}\varpi = X_{\eta} \frac{\Delta\varphi}{D\Delta\varphi} \, ,
\label{SaFNIdWylfunc}
\end{align}
we obtain
\begin{align}
{\hat\delta}_{\eta}R \overset{\mathcal{I}}{=} 2(\Delta+\kappa)X_{\eta} \, .
\label{SaFNIdetaR}
\end{align}
Similar to that of section~\pageref{sec4}, the transformation $\hat{\delta}_{\eta}$ violates not only boundary condition \ref{FNIBC1} but also boundary condition \ref{FNIBC2}. The  diffeomorphism guaranteeing non-violation of boundary condition \ref{FNIBC1} is easily obtained by setting $\delta\varphi\rightarrow\hat{\delta}_{\eta}\varphi$ in equation \ref{FNIDiffeElem}. We use this diffeomorphism to ensure that the second boundary condition is not violated,
\begin{align}
\hat{\delta}_{\eta}(\Delta R) - \frac{\Delta\hat{\delta}_{\eta}\varphi}{{D}\Delta\varphi}D(\Delta R) \overset{\mathcal{I}}{=} 0.
\label{SaFNIConsvBC2}
\end{align}
A straightforward calculation gives
\begin{align}
X_{\eta} \overset{\mathcal{I}}{=} -\frac{1}{2}\frac{d^{2}V}{d\varphi^{2}}\eta.
\label{SaFNICondnXet}
\end{align}
While this constrains $X_{\eta}$ at $\mathcal{I}^{+}$, it remains applicable within the bulk. This constraint induces a correspondingly small Weyl transformation in $\hat{\delta}_{\eta}$, including $\eta \Delta\varphi$, provided that $\eta$ falls off rapidly away from future null infinity. Consequently, to ensure the invariance of action \ref{gsIaction} under $\hat{\delta}_{\eta}$, it suffices to verify the shift transformation. This verification can be shown by
\begin{align}
{\hat\delta}_{\eta}I &= \dots + \frac{1}{16\pi G}\int_M \left( R + \frac{dV}{d\varphi}\right)\hat{\delta}_{\eta} \varphi \epsilon \, , \nonumber\\
&= \dots -\frac{1}{16\pi G}\int_M \left( \Delta R + \frac{d^{2}V}{d\varphi^{2}}\Delta\varphi \right)\eta \epsilon \, ,
\label{SaFNIWSHaction}
\end{align}
where dots are terms including $\eta\Delta\varphi$ and its derivatives. Plainly, the action is invariant under ${\hat\delta}_{\eta}$, provided that $\eta$ is small enough within the bulk. Now, it is clear why ${\hat\delta}_{\eta}$ is an approximate symmetry.

The equations of motion are obviously preserved by diffeomorphisms, but, their behavior under approximate symmetry is less obvious. Since assumed that $\eta\approx 0$ within the bulk, it suffices to check that the equations of motion are preserved by ${\hat\delta}_{\eta}$ at $\mathcal{I}^{+}$. They read
\begin{subequations}
\begin{align}
g^{ab}{\hat\delta}_\eta E_{ab} &\overset{\mathcal{I}}{=} \left(R + \frac{dV}{d\varphi}\right)(\Delta+\kappa)\eta \, , \label{SaFNIgE}\\
\ell^a\ell^b{\hat\delta}_\eta E_{ab} &\overset{\mathcal{I}}{=} \frac{1}{2}{D}\left(R + \frac{dV}{d\varphi}\right)\eta \, , \label{SaFNIllE}\\
{\hat\delta}_\eta\left( R + \frac{dV}{d\varphi}\right) &\overset{\mathcal{I}}{=} {\hat\delta}_\eta R + \frac{d^2V}{d\varphi^2}{\hat\delta}_\eta\varphi \, , \label{SaFNIdE}\\
n^an^b{\hat\delta}_\eta T_{ab} &\overset{\mathcal{I}}{=} \frac{1}{8\pi G} (\Delta-\kappa)\Delta(\Delta+\kappa)\eta \, , \label{SaFNICFT}
\end{align}
\end{subequations}
While the last one is an anomaly, not preservation, it provides a nice hint suggesting the existence of a (deformed) CFT at $\mathcal{I}^{+}$.

\section{\label{sec6} $\text{BMS}_{3}$ algebra}

The diffeomorphisms $\delta_{\xi}$ and the approximate symmetry ${\hat\delta}_{\eta}$ form a $\text{BMS}_{3}$ algebra at future null infinity, they read
\begin{alignat}{3}
&[\delta_{\xi_1}, \delta_{\xi_2}]  \overset{\mathcal{I}}{=} \delta_{\xi_{12}} \qquad
  && \hbox{with}\ \ \xi_{12} = (\xi_1\Delta\xi_2 - \xi_2\Delta\xi_1) \, , \nonumber\\
&[{\hat\delta}_{\eta_1}, {\hat\delta}_{\eta_2}]  \overset{\mathcal{I}}{=} 0 \, , \label{BMSAlBras}\\
&[\delta_{\xi_1}, {\hat\delta}_{\eta_2}]  \overset{\mathcal{I}}{=} {\hat\delta}_{\eta_{12}} \qquad
  && \hbox{with}\ \ \eta_{12} = - (\xi_1\Delta\eta_2 - \eta_2\Delta\xi_1) \, . \nonumber
\end{alignat}
Using the symplectic two-form \ref{FNIOmegI}, one may find the canonical generators corresponding to $\delta_{\xi}$ and ${\hat\delta}_{\eta}$ which satisfy the Hamilton's equation \ref{cpsm-dH}
\begin{subequations}
\begin{align}
\delta L[\xi] &= \frac{1}{8\pi G}\int_{\mathcal{I}^{+}}\left(\delta\varphi\,\delta_\xi\kappa
    - \delta_\xi\varphi\,\delta\kappa\right) \ell_\mu \nonumber\\
    &= \frac{1}{8\pi G}\int_{\mathcal{I}^{+}}\left( \delta\varphi\,\Delta(\Delta+\kappa)\xi
    - \xi \Delta\varphi\,\delta\kappa \right) \ell_\mu  \, , \label{BMSAlinftsmlCanoL} \\
\delta M[\eta] &= \frac{1}{8\pi G}\int_{\mathcal{I}^{+}}\left( \delta\varphi\,{\hat\delta}_\eta\kappa
    - {\hat\delta_\eta}\varphi\,\delta\kappa \right) \ell_\mu \nonumber\\
    &= \frac{1}{8\pi G}\int_{\mathcal{I}^{+}}\left(
    - \hat{\delta}_{\eta}\omega\, \Delta\delta\varphi -\delta\kappa(\Delta+\kappa)\eta \right)\ell_\mu \, .
\label{BMSAlinftsmlCanoM}
\end{align}
\end{subequations}
To guarantee the integrability of the infinitesimal canonical generators, it is crucial to consider the procedure \ref{FNIRealH}. The compatible canonical generators are therefore given by
\begin{subequations}
\begin{align}
L[\xi] &= \frac{1}{8\pi G}\int_{\mathcal{I}^{+}}\left(\xi \Delta^2\varphi - \kappa\xi \Delta\varphi\right)\ell_\mu \, ,
\label{BMSAlCanoL} \\
M[\eta] &= \frac{1}{8\pi G}\int_{\mathcal{I}^{+}} \left(\Delta\kappa - \frac{1}{2}\kappa^2\right)\eta \ell_\mu \, .
\label{BMSAlCanoM}
\end{align}
\end{subequations}
Applying equation \ref{cpsm-PBra} yields the following Poisson brackets
\begin{align}
&\left\{L[\xi_1],L[\xi_2]\right\} \overset{\mathcal{I}}{=}  L[\xi_{12}] \, , \nonumber\\
&\left\{M[\eta_1],M[\eta_2]\right\} \overset{\mathcal{I}}{=}  0 \, , \label{BMSAlPoBra}\\
&\left\{L[\xi_1],M[\eta_2]\right\} \overset{\mathcal{I}}{=} \nonumber\\
& -M[\eta_{12}]
   - \frac{1}{16\pi G}\int_{\mathcal{I}^{+}} \left(\Delta\xi_1 \Delta^2\eta_2 - \Delta\eta_2 \Delta^2\xi_1\right)\ell_\mu \, ,
\nonumber
\end{align}
where $\xi_{12}$ and $\eta_{12}$ were given in equation \ref{BMSAlBras}. This is in agreement with $\text{BMS}_{3}$ algebra \ref{BMSAlBras}, however, a central term interestingly stands out here.

\section{\label{sec7} From moding to counting}

The Poisson brackets \ref{BMSAlPoBra} may be realized as a quantum $\text{BMS}_{3}$ algebra by applying the usual rule for substituting from Poisson brackets to commutators $\{\cdot,\cdot\}\rightarrow-\frac{i}{\hbar}\left[\cdot,\cdot\right]$. To this end, we need to perform a moding procedure on the canonical generators. The standard mode expansion yields $e^{im\kappa u}$ for modes in the presence of a black hole, where $u$ is the retarded time along future null infinity. In a coordinate system like Gaussian null, $\kappa$ would be $\sim\mathcal{O}(1/r^{2})$ at $\mathcal{I}^{+}$, so modes of the form $e^{im\kappa u}$ are compatible with soft gravitons in higher dimensions. Let us take $\psi=\kappa u$ as a phase with $D\psi\overset{\mathcal{I}}{=}0$ and normalize the retarded time so that $\Delta u=1$, then, using the identity \ref{gsIdentc}, we get $d\psi\overset{\mathcal{I}}{=}-\kappa\ell_{\mu}$ with $\Delta\kappa\ll\kappa$ assumption. Defining suitable modes as
\begin{align}
\xi_{m} \overset{\mathcal{I}}{=} \frac{1}{\kappa} e^{im\psi} \quad \text{and} \quad \eta_{n} \overset{\mathcal{I}}{=} \frac{1}{\kappa} e^{in\psi}, \label{FMTCmodes}
\end{align}
and applying the substitution mentioned above, we obtain the quantum $\text{BMS}_{3}$ algebra
\begin{align}
&[L_{m},L_{n}]= (m-n)L_{m+n} \, , \nonumber\\
&[M_{m},M_{n}] = 0 \, , \label{FMTCqbms3}\\
&[L_{m},M_{n}]
   = (m-n)M_{m+n} + c_{\scriptscriptstyle LM} m(m^2-1)\delta_{m+n,0} \, ,
\nonumber
\end{align}
where the central charge is given by performing the integration in \ref{BMSAlPoBra}, it reads
\begin{align}
 c_{\scriptscriptstyle LM}=\frac{1}{4G} \, . \label{FMTCmodes}
\end{align}
Generally, the BMS$_{3}$ algebra can have two different nontrivial central extensions. The dimensionless central charge of the diffeomorphism algebra vanishes in our study because we are dealing with Einstein-dilaton gravity dimensionally reduced from pure Einstein gravity. There are some studies~\cite{Bagchi:2012yk} suggesting that the dimensionless central charge can be nontrivial when one deals with the theories, namely topological massive gravity, in which the parity is broken~\cite{Oblak:2017ptc}. The dimensionful central charge \eqref{FMTCmodes} was already realized for the asymptotic symmetries of asymptotically flat spacetimes at null infinity in three dimensions~\cite{Barnich:2006av}. Customarily, when dealing with theories like AdS gravity, which has a length scale, we can exploit this parameter to make the central charge dimensionless. However, in our study of asymptotically flat gravity, this is difficult since there is no such length scale. Nonetheless, it might be possible to introduce an arbitrary parameter and rescale $\eta$ to make the central charge dimensionless. In such a scenario, one would need to justify the introduced parameter with physical motivations~\cite{Carlip:1998wz}.

From the dual field theory side, the algebra \ref{FMTCqbms3} is equivalent to a Galilean CFT living at $\mathcal{I}^{+}$. In~\cite{Bagchi:2012xr}, it was shown that a Cardy-like formula also works for such a Galilean CFT, and so, gives the microscopic entropy by counting microstates. As usual, we have
\begin{align}
S_{\text{mic}} = 2\pi h_{\scriptscriptstyle L}\sqrt{\frac{c_{\scriptscriptstyle LM}}{2h_{\scriptscriptstyle M}}}.
\label{FMTCCardy}
\end{align}
We shall determine the eigenvalues, $h_{\scriptscriptstyle L}$ and $h_{\scriptscriptstyle M}$, of the zero-modes of $L$ and $M$. Applying the boundary condition \ref{FNIBC1} to \ref{BMSAlCanoL} and \ref{BMSAlCanoM} gives
\begin{subequations}
\begin{align}
h_{\scriptscriptstyle L} = L[\xi_0] &= -\frac{\left(\kappa\varphi^{N-2}\right)_{\mathcal{I}^{+}}^{\frac{N-2}{N-3}}\left(\frac{a}{N-3}\right)^{\frac{N-2}{N-3}}}{16\pi^{2} G}\int_{\mathcal{I}^{+}} \kappa \ell_\mu \nonumber \\
&= \frac{\left(\kappa\varphi^{N-2}\right)_{\mathcal{I}^{+}}^{\frac{N-2}{N-3}}\left(\frac{a}{N-3}\right)^{\frac{N-2}{N-3}}}{16\pi^{2} G}\int_{\mathcal{I}^{+}} d\psi \nonumber \\
& = \frac{\left(\kappa\varphi^{N-2}\right)_{\mathcal{I}^{+}}^{\frac{N-2}{N-3}}\left(\frac{a}{N-3}\right)^{\frac{N-2}{N-3}}}{8\pi G}, \label{FMTChL} \\
h_{\scriptscriptstyle M} = M[\eta_0] &= -\frac{1}{16\pi G}\int_{\mathcal{I}^{+}} \kappa \ell_\mu \nonumber \\
   &= \frac{1}{16\pi G}\int_{\mathcal{I}^{+}} d\psi = \frac{1}{8G}, \label{FMTChM}
\end{align}
\end{subequations}
where $(\kappa\varphi^{N-2}/(N-3))_{\infty}$ gives the mass of the black hole, as it can be easily checked for some black hole solutions. Another verification is to consider the time translation $\xi^{\mu}=n^{\mu}$, we get then
\begin{align}
\delta E &= \frac{1}{8\pi G}\int_{\mathcal{I}^{+}} \left( \delta\varphi\Delta\kappa - \Delta\varphi \delta\kappa \right) \ell_\mu \, , \nonumber\\
E &= \frac{1}{8\pi G}\int_{\mathcal{I}^{+}} \left( \varphi\Delta\kappa \right) \ell_\mu \, . \label{FMTCEformula}
\end{align}
Imposing $\Delta\kappa\overset{\mathcal{I}}{=}-\kappa^{2}\varphi^{N-3}/2\pi(N-3)$ as a new boundary condition leads to
\begin{align}
E &= \frac{(\kappa\varphi^{N-2})_{\mathcal{I}^{+}}}{8\pi G (N-3)} = \frac{(\kappa\varphi^{N-2})_{i^{0}}}{8\pi G (N-3)} \, , \label{FMTCEMass}
\end{align}
where the last equality is obtained by the integration by parts. This is the well-known relation $E=Mass/8\pi G$.

There is one more subtlety to be addressed. The derivation of equation \ref{BMSAlinftsmlCanoL} relies on integration by parts, which incorporates a boundary term,
\begin{align}
\delta L[\xi]\Bigl|_{i^0} = \frac{1}{8\pi G}\left(\xi \Delta\delta\varphi
   - \delta\varphi\,(\Delta+\kappa)\xi\,\right)\Bigl|_{i^{0}} \, .
\label{FMTCsubtletydL}
\end{align}
As discussed in section~\pageref{sec4}, we have $\Delta\delta\varphi\overset{i^{0}}{=}0$. Taking $\kappa$ to be constant at $i^{0}$, we may write
\begin{align}
\delta L[\xi]\Bigl|_{i^0} = -\frac{1}{8\pi G} \delta\left(\varphi\,(\Delta+\kappa)\xi\,\right)\Bigl|_{i^{0}} \, .
\label{FMTCdivergPhiterm}
\end{align}
Clearly, the dilaton field is divergent at $i^{0}$ even for the zero-mode of $L$. As argued in section \pageref{sec2}, we recall that the term involving the spatial boundary may contribute to calculations. The spatial boundary term is a total derivative $\Theta\Bigl|_{i^{0}}=dC$, therefore, one may write $\delta L[\xi]\Bigl|_{i^{0}}=\int_{i^0}\delta\Theta=\int_{i^0} d(\delta C)=\delta C$. Such a contribution was worked out in~\cite{Harlow:2019yfa}, for our study it reads
\begin{align}
\delta L[\xi]\Bigl|_{i^0} = \delta C &= -\frac{1}{8\pi G} \delta\left(\varphi\,\gamma^{\mu\nu} \hat{n}^{\alpha}\delta_{\xi} g_{\nu\alpha}\ell_{\mu}\,\right)\Bigl|_{i^{0}} \,  \nonumber\\
                                     &= -\frac{1}{8\pi G} \delta\left(\varphi\,\gamma^{\mu\nu} \hat{n}^{\alpha}g_{\nu\alpha}(\Delta+\kappa)\xi\ell_{\mu}\,\right)\Bigl|_{i^{0}} \nonumber\\
                                     &= \frac{1}{8\pi G} \delta\left(\varphi\,(\Delta+\kappa)\xi\,\right)\Bigl|_{i^{0}}, \label{FMTCCterm}
\end{align}
where $\hat{n}^{\alpha}$$=($$n^{\alpha}$$-$$\ell^{\alpha}$)$/$$\sqrt{2}$ and $\gamma^{\mu\nu}$$=$$g^{\mu\nu}$$-$$\hat{n}^{\mu}$$\hat{n}^{\nu}$. As seen, the spatial boundary contribution cancels the divergent term \ref{FMTCdivergPhiterm}.

With all concerns addressed, we are now able to calculate the entropy by the Cardy formula. Plugging \ref{FMTCmodes}, \ref{FMTChL}, and \ref{FMTChM} into \ref{FMTCCardy} yields
\begin{align}
S_{\text{mic}}=\frac{\left(\kappa\varphi^{N-2}\right)_{i^{0}}^{\frac{N-2}{N-3}}\left(\frac{a}{N-3}\right)^{\frac{N-2}{N-3}}}{4 G}=\frac{\varphi^{N-2}_{\mathcal{B}}}{4G}=S_{2D}  \label{GDentropy}
\end{align}
which is exactly the Bekenstein-Hawking entropy for 2D dilaton gravity. Our result is in agreement with Carlip's result~\cite{Carlip:2017xne,Carlip:2019dbu} which was done by considering the dual field theory at the future horizon. There are multiple ways to perform the dimensional reduction to two dimensions\footnote{See~\cite{Cavaglia:1998xj} and references therein.}, as all dimensionally reduced 2D solutions are conformally equivalent. A considerable approach is to maintain the ``r-t'' plane of the 2D metric identical to the original higher dimensional metric. This ensures that the 2D surface gravity is precisely the same as the higher dimensional surface gravity\footnote{While the other reduction approaches yield correct surface gravity scalars with the same value at the horizon, they may not have the required physical parameters, such as the mass of the black hole, in their expressions}. In this case, the dilaton field would have the unit of length, exemplified by $\varphi=r$ for Schwarzschild or RN black holes, where $r$ is the usual radial coordinate. The field redefinition $\varphi^{N-2}=\phi$ leads to the standard entropy found in the literature, $S=\phi_{\mathcal{B}}/4G$.

Note that the 2D action \ref{gsIaction} was given up to integration over ``$N-2$'' transverse coordinates ``$y$''. This implies that for $N$-dimensional theory, as argued by Carlip in~\cite{Carlip:2019dbu}, one shall add all individual entropies by performing the integration over transverse coordinates. This is in accord with the nature of extensive entropies. The result is then
\begin{align}
S_{N} = \int d^{N-2}y\,S_{2D} =  \frac{A_{hor}}{4G} \, .
\label{GDNentropy}
\end{align}
where $A_{hor}$ is the horizon area. The boundary condition \ref{FNIBC1} and subsequently the method of the ``black hole microstate counting'' presented in this work offer some phenomenological insights, suggesting that the dual (deformed) CFT depends on the specific black hole being investigated. The main distinction between counting black hole microstates at the horizon and at the future null infinity lies in their differing boundary conditions. It is worth mentioning that the presence of the factor ``$a$'' in the boundary condition \ref{FNIBC1} does not contradict the universality of entropy, as the entropy of any black hole solution adheres to the area law. Interestingly, the necessity of a constant value to define the behavior of the dilaton field at the asymptotic region was previously noted in a completely different context~\cite{Engelsoy:2016xyb,Almheiri:2014cka}. However, in this study, this constant serves not only as a boundary condition but also as a crucial parameter differentiating various black holes as observed by an asymptotic null observer.

For a null observer at the asymptotic region, the dominant wave sector is the s-wave. This is the reason that the present calculations work for any type of asymptotically flat black hole solutions as far as they are considered at the asymptotic region\footnote{Clearly, the argument is also true for asymptotically (A)dS solutions, but, those are not the subject of the present study.}. Nonetheless, the s-wave reduction, and therefore the present study, may be still applicable to finite distances from the horizon in the case where the solution is invariant under the rotation of the $S^{2}$\footnote{See~\cite{Nayak:2018qej} for a detailed discussion on s-wave reduction and spherically symmetric solutions.}. In such cases, the quadrupole moments of the gravitational field should also be taken into account. Such investigations will be arduous, yet they may illuminate the nature of quantum gravity and its underlying degrees of freedom.

Although this study focused exclusively on asymptotically flat black holes, the boundary condition \ref{FNIBC1} does not admit three-dimensional asymptotically flat black hole solutions. However, this is not a major concern for two reasons. Firstly, it is widely recognized that such black hole solutions are uncommon in 2+1 pure Einstein gravity~\cite{Alkac:2016xlr}. Consequently, the literature often explores theories beyond pure Einstein gravity, such as new massive gravity in three dimensions~\cite{Bergshoeff:2009hq}. Secondly, if one begins, for example, with four-dimensional pure Einstein gravity and then performs dimensional reduction to obtain a three-dimensional black hole solution, the presented procedure cannot be applied to study the resulting three-dimensional solution by further dimensional reduction to two dimensions. This is because, in such a case, the dimensionally reduced 2+1 solution is no longer a pure Einstein gravity but rather a 2+1 dilaton-gravity system. In summary, a thorough examination of such dimensional reductions is necessary, and it is anticipated that they adhere to boundary conditions specific to themselves.

\section{\label{sec8} Conclusions}

We have demonstrated that counting black hole microstates may be achieved at future null infinity. Two essential steps are first to perform a dimensional reduction from higher dimensional theories to two dimensions and second to impose some boundary conditions to handle the asymptotic behavior of observable charges while avoiding non-integrability and destroying information. Holography at infinity has a rich phenomenological aspect because it is expected that someday the black hole information will be extractable through future detectors. As argued in section~\pageref{sec1} , our physical justification to consider black hole holography at future null infinity is that there are possibly soft gravitons that bring the information from the past horizon. Our work is supported by a recent study by Laddhaa et al.~\cite{Laddha:2020kvp} in which they found that all information about massless excitations is readable through a tiny region near the very beginning edge of future null infinity.

Let us discuss our boundary conditions. In adherence to the principles of canonical formalism, the selection of boundary condition \ref{FNIBC1} has been made with the specific intent of including solely the canonical variables, thereby excluding extraneous quantities. If one accepts that it is safe to choose the boundary condition such that it includes the quantities like horizon radius $r_{\text{hor}}$, then, one may write the boundary condition \ref{FNIBC1} as $\Delta\varphi\overset{\mathcal{I}}{=}(\kappa/2\pi)r_{\text{hor}}^{N-2}$ and $\Delta\Delta\varphi\overset{\mathcal{I}}{=}0$. This boundary condition leads to
\begin{align}
S_{\text{mic}}=\frac{r_{\text{hor}}^{N-2}}{4 G} \label{FGDentropy}
\end{align}
which is actually the same as \ref{GDentropy}, but in a more manifest way. Moreover, this shows us something interesting: it may be possible to consider boundary conditions like $\Delta\varphi\overset{\mathcal{I}}{=}(\kappa/2\pi)(r_{+}^{N-2}+r_{-}^{N-2})$ or even $\Delta\varphi\overset{\mathcal{I}}{=}(\kappa/2\pi)(r_{+}^{N-2}+r_{-}^{N-2}+\dots)$, where $r_{+}$, $r_{-}$, and dots are respectively outer horizon radius, inner horizon radius, and possible correlations between them. These results not only provide a holographic interpretation for non-extremal black holes but also interestingly imply that the detection of soft gravitons if it occurs someday, would shed light on non-equilibrium black hole thermodynamics.

As final remarks: we argue that although it seems detecting gravitons, of course including soft ones, is a long way off, it may be possible to look for the dual field theory observables living on the holographic plate at infinity, as it is endorsed by the present calculations. Another argument is that one may perform the same procedure to count black hole microstates at the past horizon as it was done by Carlip at the future horizon. Interestingly, the microscopic entropy can be calculated at every single side of the Penrose diagram, except past null infinity which is like the lack of black hole information in the exterior region which is the meaning of entropy itself!

For future studies: one may consider asymptotically AdS or dS spacetimes. Nonetheless, such studies have a long history in asymptotically AdS spacetimes~\cite{Brown:1986nw}, they will be challenging in the concept of Newman-Penrose formalism and relevant Cauchy surface at asymptotic region. One may also consider charges calculation in Newman-Penrose formalism without a dimensional reduction, a start has been made in~\cite{Liu:2022uox} for the near-horizon. Last, but not least, is to investigate possible connections between our work and recent studies on null boundary phase space like~\cite{Chandrasekaran:2018aop,Adami:2021nnf}. Recent work by Ashtekar et al.~\cite{Ashtekar:2024bpi} argues that the information is encoded in the intrinsic connection at $\mathcal{I}^{+}$, which is consistent with our calculations.

\section{\label{sec9} Acknowledgments}
The author would like to express sincere gratitude to the anonymous referee for their insightful comments and constructive suggestions, which significantly improved the clarity and quality of this work. The author would like to express warm gratitude to Steve Carlip for his initial encouragement and valuable comments, which provided the impetus for this research. The author extends sincere thanks to Shahin Sheikh-Jabbari for his valuable comments and support. The author also would like to thank Bahman Khanpour, Ali Farokhtabar, Marjan Kioumarsipour, Hamed Adami, and especially Ali Seraj for valuable feedbacks on the initial draft. This work is based upon research funded by Iran National Science Foundation
(INSF) under project No. 4031681.

\bibliographystyle{aipnum4-2}
\bibliography{Shajiee2024}

\end{document}